\documentclass[conference]{IEEEtran}
\IEEEoverridecommandlockouts
\usepackage{cite}
\usepackage{amsmath,amssymb,amsfonts}
\usepackage{algorithmic}
\usepackage{graphicx}
\usepackage{textcomp}
\usepackage{xcolor}
\usepackage{booktabs}       
\usepackage{hyperref}       
\usepackage{etoolbox}

\def\BibTeX{{\rm B\kern-.05em{\sc i\kern-.025em b}\kern-.08em
    T\kern-.1667em\lower.7ex\hbox{E}\kern-.125emX}}

\begin{document}

\title{Interpretable Dual-Filter Fuzzy Neural Networks for Affective Brain-Computer Interfaces}

\author{
\IEEEauthorblockN{Xiaowei Jiang, Yanan Chen, Nikhil Ranjan Pal, Yu-Cheng Chang, Yunkai Yang, Thomas Do, Chin-Teng Lin\textsuperscript{*}}

\thanks{Xiaowei Jiang, Yu-Cheng Chang,  Thomas Do, and Chin-Teng Lin are with the GrapheneX-UTS Human-centric AI Centre, Australian AI Institute, School of Computer Science, Faculty of Engineering and Information Technology, University of Technology Sydney.}
\thanks{Yanan Chen and Yunkai Yang are with the Institute of Psychology and Behavior, Henan University.}
\thanks{Nikhil Ranjan Pal is with the Indian Statistical Institute, Kolkata, India}
\thanks{
This work was supported in part by the Australian Research Council (ARC) under discovery grant DP220100803 and DP250103612 and ITRH grant IH240100016, Australian National Health and Medical Research Council (NHMRC) Ideas Grant APP2021183, and the UTS Human-Centric AI Centre funding sponsored by GrapheneX (2023-2031). The research was also sponsored in part by the Humanities and Social Sciences project of Ministry of Education (22YJCZH021) and the Institute of Psychology, Chinese Academy of Sciences (Grant No. GJ202007).}
\thanks{\textsuperscript{*}Corresponding author: Chin-Teng Lin. Email: chin-teng.lin@uts.edu.au}
}

\maketitle
\begin{abstract}
Fuzzy logic provides a robust framework for enhancing explainability, particularly in domains requiring the interpretation of complex and ambiguous signals, such as brain-computer interface (BCI) systems. Despite significant advances in deep learning, interpreting human emotions remains a formidable challenge. In this work, we present \textit{iFuzzyAffectDuo}, a novel computational model that integrates a dual-filter fuzzy neural network architecture for improved detection and interpretation of emotional states from neuroimaging data. The model introduces a new membership function (MF) based on the Laplace distribution, achieving superior accuracy and interpretability compared to traditional approaches. By refining the extraction of neural signals associated with specific emotions, \textit{iFuzzyAffectDuo} offers a human-understandable framework that unravels the underlying decision-making processes. We validate our approach across three neuroimaging datasets using functional Near-Infrared Spectroscopy (fNIRS) and Electroencephalography (EEG), demonstrating its potential to advance affective computing. These findings open new pathways for understanding the neural basis of emotions and their application in enhancing human-computer interaction.
\end{abstract}


\section{Introduction}

\begin{figure*}[ht!]
    \centering
    \includegraphics[width=0.8\linewidth]{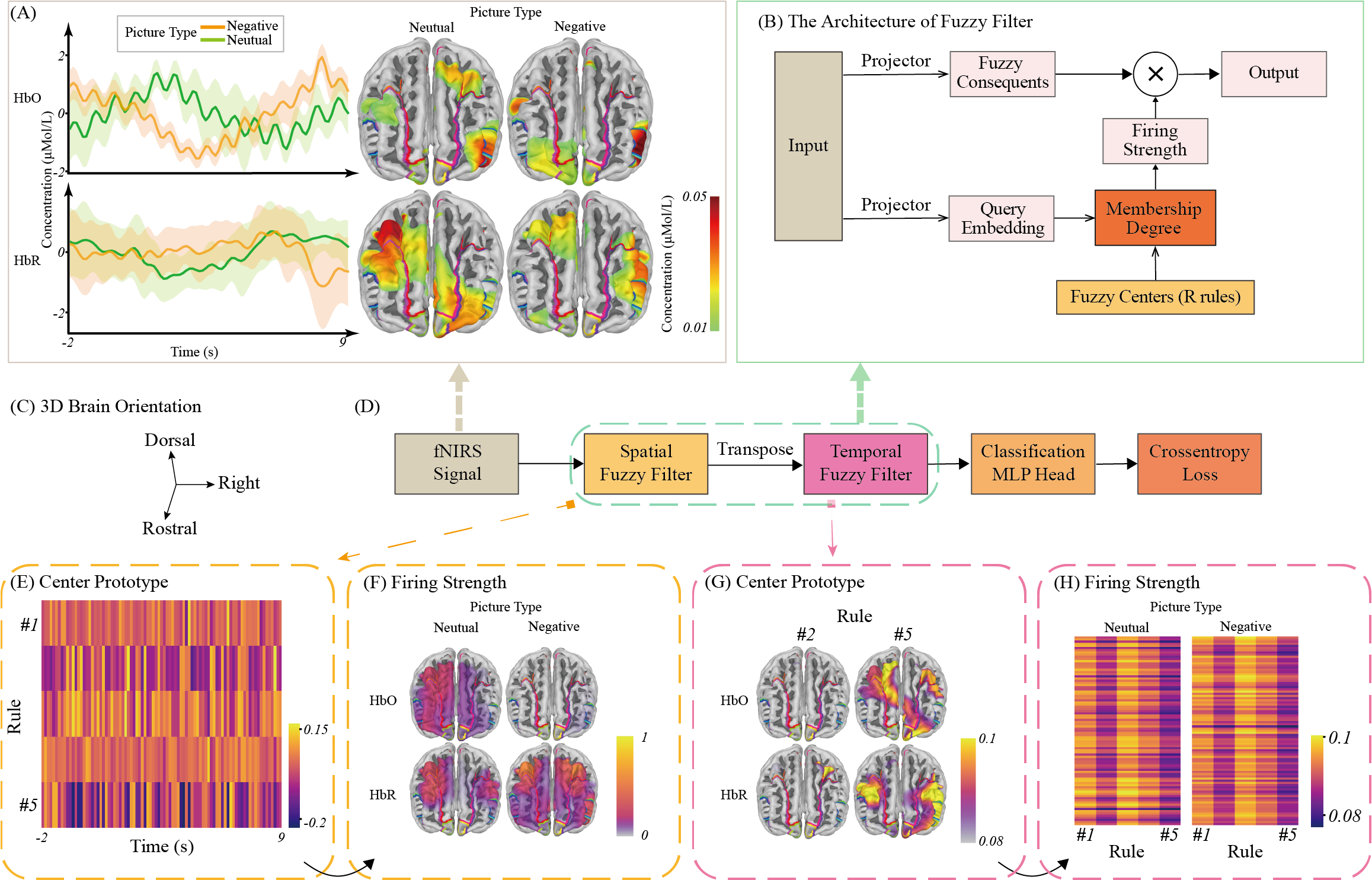}
    \caption{\textbf{Overview of the iFuzzyAffectDuo model and related components.} 
    \textbf{(A)} Example of the fNIRS signal recorded during exposure to affective stimuli, showcasing both negative and neutral responses.
    \textbf{(B)} Architecture of the fuzzy filter employed in the model.
    \textbf{(C)} Brain orientation for all 3D plots used in the analysis.
    \textbf{(D)} Structural layout of the proposed iFuzzyAffectDuo model.
    \textbf{(E)} Center prototypes of the spatial filter.
    \textbf{(F)} Firing Strengths of the spatial filter.
    \textbf{(G)} Center prototypes of the temporal filter.
    \textbf{(H)} Firing Strengths of the temporal filter.}
    \label{fig:model_arch}
\end{figure*}

\IEEEPARstart{E}{motions} are fundamental to human existence, shaping diverse aspects of life and evolving over millions of years to form the cornerstone of social intelligence and group cohesion~\cite{EMOTION1884JAMES, SolvingLisa, BriefHistoryofHumanSociety2002Douglas}. The ability to recognize and respond to emotional cues across varying contexts—such as how anger can enhance the detection of fear and vice versa—is crucial for survival and social functioning~\cite{Mumenthaler2012SocialAI}. Advances in neuroimaging have illuminated the neural pathways involved in processing affective stimuli, emphasizing the roles of appetitive and defensive systems rooted in primitive neural circuits~\cite{langEmotionMotivationAttention2000}. These systems coordinate responses ranging from attentional shifts to physiological changes, highlighting their critical role in survival and the dynamic interplay between physiological states and psychological experiences~\cite{Volchan2003EmotionalSS, langLookingPicturesAffective1993, Toga2015EmotionalE, Porges2008ReciprocalIB}. Despite significant progress, mapping specific neural circuits to distinct emotions remains challenging, with only probabilistic patterns identified thus far~\cite{Toga2015EmotionalE}. Understanding how the human brain processes emotions remains a priority, spurring the development of Affective Computing (AC) systems, particularly Affective Brain-Computer Interfaces (aBCIs). These systems aim to bridge the gap between human emotions and machine understanding by recognizing, interpreting, processing, and simulating emotional states across various modalities~\cite{khareEmotionRecognitionArtificial2024a, rahmanRecognitionHumanEmotions2021a, ACReview2005Tao, Picard1997AffectiveC}.

The aBCIs system integrate data from diverse sources, including behavioral cues~\cite{Cambria2017}, peripheral nervous system (PNS) signals such as heart rate variability (HRV)\cite{7574455}, and central nervous system signals, including EEG~\cite{xiao4DAttentionbasedNeural2022} and fNIRS~\cite{spapeNEMODatabaseEmotion2024}. These approaches offer a comprehensive perspective on the neural and physiological underpinnings of emotions~\cite{peiAffectiveComputingRecent2024}. Recent advancements in aBCIs have demonstrated the effectiveness of Convolutional Neural Networks (ConvNets) for end-to-end emotional decoding from EEG data. For instance, EEGNet has set new benchmarks in performance~\cite{Lin2023ConvolutionalNN}, while the NLSTM model has showcased exceptional sensitivity and specificity across diverse EEG datasets, emphasizing its robustness and generalization capabilities in both controlled and real-world scenarios~\cite{doi:10.1142/S0129065720500197}. Similarly, DBjNet has achieved remarkable decoding accuracy in distinguishing between negative and neutral emotions using fNIRS data~\cite{siCrossSubjectEmotionRecognition2023}. However, despite these advancements, the interpretability of these models remains a significant challenge, limiting the understanding of their decision-making processes and hindering their broader application in real-world contexts.

Addressing this limitation, Fuzzy Neural Networks (FNNs) combine the adaptability of neural networks with the interpretability of fuzzy logic systems. FNNs leverage fuzzy rules and membership functions (MFs) to process inputs, providing an intuitive framework that contrasts sharply with the opacity of conventional deep learning models~\cite{lin1996neural, 10183374,106218}. This transparency facilitates a clear understanding of how inputs are transformed into outputs, enhancing users' trust and interaction in aBCI systems~\cite{jiangFuzzybasedApproachPredict2024, jiangIFuzzyTLInterpretableFuzzy2024, decampossouzaInterpretableEvolvingFuzzy2021}. FNNs have been successfully applied to pattern classification tasks~\cite{PEDRYCZ19931} and extended to adaptively learn spatiotemporal knowledge through advanced architectures like Evolving Fuzzy Neural Networks (EFuNNs)~\cite{kasabovEvolvingFuzzyNeural2001} and the interpretable fuzzy transfer learning model (iFuzzyTL)~\cite{jiangIFuzzyTLInterpretableFuzzy2024}.

Building on this foundation, we introduce the Interpretable Dual-Filter Fuzzy Rule-Based Model for aBCIs (iFuzzyAffectDuo), an evolution of the iFuzzyTL model originally designed for transfer learning in  (Steady-state visually evoked potential) SSVEP-based EEG tasks. While iFuzzyTL demonstrate strong domain adaptation capabilities, it faces limitations in interpretability and complex pattern recognition due to the relatively simple neural patterns associated with SSVEP~\cite{jiangIFuzzyTLInterpretableFuzzy2024}. iFuzzyAffectDuo overcomes these limitations by incorporating a dual-filter structure that combines spatial and temporal filters, inspired by EEGNet, to enhance feature extraction and pattern recognition~\cite{lawhernEEGNetCompactConvolutional2018}, and transiting from Gaussian to Modified Laplace MFs, significantly improving performance.

Furthermore, iFuzzyAffectDuo integrates a fuzzy attention mechanism, enabling the model to generalize central fuzzy rules while capturing domain-specific spatiotemporal dependencies in brain signals. This novel architecture supports robust feature extraction and excels in emotion classification tasks, making it particularly effective for aBCI applications~\cite{9762054, CHEN2023105312}. To validate its performance, we evaluate iFuzzyAffectDuo on three datasets: two fNIRS datasets (Picture Recognition and Picture Empathy) and one EEG dataset (FACED) for emotion recognition tasks. Experimental results demonstrate that iFuzzyAffectDuo achieves state-of-the-art accuracy and interpretability, effectively capturing affective neural patterns and advancing the performance of aBCI systems across both fNIRS and EEG modalities.

\section{Methodology}
\label{sec:methodology}

\subsection{Fuzzy Inference Systems}
Fuzzy Inference Systems (FISs) are widely used to model uncertainty and imprecision in various domains.  A class of FIS, realized using neural architectures, is known as Fuzzy Neural networks  (FNNs)~\cite{lin1996neural}, which generally use gradient descent optimization for training. This system uses the concept of membership, which quantifies the degree to which an element \(x\) belongs to a fuzzy set characterized by a membership function \(A(x)\). One prominent approach to fuzzy modeling is the Takagi-Sugeno-Kang (TSK) model~\cite{shihabudheen2018recent}, which utilizes a set of IF-THEN rules to define the relationship between inputs and outputs. For a Zero-th order TSK fuzzy  system, the rules are expressed as:
\begin{equation}
    \mathrm{If}\ x_1\ \mathrm{is}\ A_{1,r}, \cdots, x_D ~\mathrm{is}~ A_{D,r} ,
\end{equation}  
\begin{equation}    
    \mathrm{then\ the\ output\ is}\ y= u_{r}; r=1, \cdots, R,
\end{equation}
where \(x_i; i=1, \cdots, D\) represent the $d$ input (linguistic) variables and $A_{i,r}; i=1, \cdots, D$ are $D$ fuzzy sets for the $r^{th}$ rule, which are defined by the membership functions $A_{i,r}(x_i); i=1, \cdots, D.$

The firing strength \(\mu_r\) of the $r_{th}$ rule  is typically computed as the product of the individual MFs, i.e., 
\begin{equation}
\label{equ:mfs_d}
\mu_r(x)=\Pi_{i=1}^D A_{i,r}(x_i)
\end{equation}
The final output of the TSK FIS is derived as:
\begin{align}
y = \sum_{j=1}^{R}{\frac{\mu_j(x)u_j}{\sum_{i=1}^{R}{\mu_i(x)}}},
\label{eq.rule_out}
\end{align}
where \(R\) is the total number of rules, and \(y\) represents the aggregated output obtained through weighted aggregation of the rule outputs. If the system has multiple outputs, each rule will have multiple consequents.

\subsection{Enhancing Sensitivity to Scale Parameter with Modified-Laplace Membership Functions}
\begin{figure}[ht]
    \centering
    \includegraphics[width=1\linewidth]{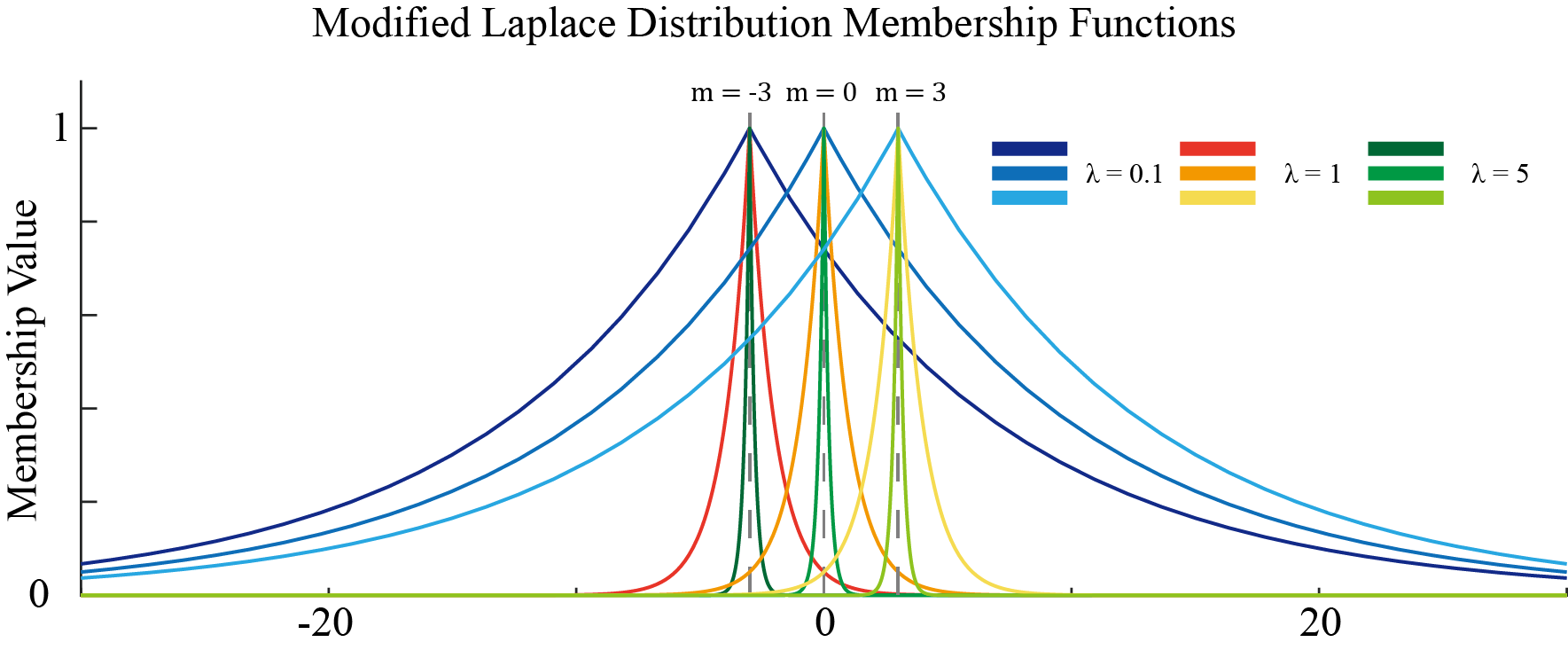}
    \caption{Modified-Laplace MFs $m$ with different center parameters $m$ and $\lambda$}
    \label{fig:mf}
\end{figure}

Traditional Gaussian MFs often exhibit broad widths, which can lead to suboptimal feature representation within models~\cite{jiangFuzzybasedApproachPredict2024}. To mitigate this issue, we introduce a novel MF inspired by the Laplace Distribution, defined by:
\begin{equation}
f(x \mid m, b) = \frac{1}{2b} e^{-\frac{|x - m|}{b}},
\end{equation}
where $m$ is the location parameter and $b$ is the scale parameter. We modify this distribution as follows to better suit our fuzzy systems as Modified-Laplace MFs: 
\begin{equation}
\mu_{ML}(x \mid m, \lambda) = e^{-\lambda |x - m|},
\end{equation}
where $\lambda_d$ is a width factor within the range of $[0, +\infty)$. Both $\lambda_d$ and $m$ are trainable parameters. The firing strength $\mu_r(x)$ in our TSK model is computed using the product of these Modified-Laplace MFs, following the eq. (\ref{equ:mfs_d}):
\begin{equation}
\mu_r(x) = \prod_{d} e^{-\lambda_d |x_{d} - m_{r,d}|},
\end{equation}
where $x_d$ denotes the $d$-th feature, $m_{r,d}$ is the center of the rule $r$'s fuzzy set for the $d$-th feature, and $\lambda_d$ dynamically adjusts the sensitivity of the MF, as shown in Fig.~\ref{fig:mf}.

The firing strength of a fuzzy rule in the TSK model, $\overline{f_{ML}}_{i,r}(\mathbf{x})$, is expressed as:
\begin{align}
\label{eq:fs}
\overline{f_{ML}}_{i,r}(\mathbf{x}) &= \frac{\mu_r(\mathbf{x})}{\sum_{j=1}^R \mu_j(\mathbf{x})} \\
&= \frac{e^{-\sum_{d=1}^D \lambda_{i,d} |x_{i,d} - m_{r,d}|}}{\sum_{j=1}^R e^{-\sum_{d=1}^D \lambda_{j,d} |x_{i,d} - m_{r,d}|}}
\end{align}

\subsection{Sensitivity Analysis of Modified-Laplace Membership Functions}

To conduct the Sensitivity Analysis, we analyze the derivative of the Gaussian MFs with respect to the width parameter \(\sigma\) and the Modified-Laplace MF with respect to the width parameter \(\lambda\) to show how sensitive each MF is to changes in its respective width parameters. 

The derivative of \(f_{\text{G}}(x \mid m,\sigma)\) with respect to \(\sigma\) is given by:
\begin{equation}
\frac{\partial \mu_{\text{G}}(x \mid m,\sigma)}{\partial \sigma} = \mu_{\text{G}}(x \mid m,\sigma) \cdot \frac{(x - m)^2}{\sigma^3}.
\end{equation}

The derivative of \(\mu_{\text{ML}}(x \mid m,\lambda)\) with respect to \(\lambda\) is given by:
\begin{equation}
\frac{\partial \mu_{\text{ML}}(x \mid m,\lambda)}{\partial \lambda} = -|x - m| \cdot \mu_{\text{ML}}(x \mid m,\lambda).
\end{equation}

The Gaussian MF is more sensitive to changes in its width parameter \(\sigma\) compared to the Modified-Laplace function's sensitivity to \(\lambda\). This higher sensitivity arises because the Gaussian derivative scales with the square of the distance from the mean, \((x - m)^2\), and inversely with \(\sigma^3\). In contrast, the ML derivative scales linearly with \(|x - m|\). As a result, the Gaussian's sensitivity increases more rapidly as the input deviates from the mean and is more affected by small changes in \(\sigma\).

\subsection{iFuzzyAffectDuo Main Structure}

The proposed model, iFuzzyAffectDuo, is composed of three principal modules, as shown in Fig.~\ref{fig:model_arch}(D). The first is the Spatial Fuzzy Filter, which is designed to capture spatial patterns within the brain data. This filter facilitates the nuanced detection of region-specific neural activity, enabling a deeper understanding of spatial dynamics associated with emotional responses. 
The second module, the Temporal Fuzzy Filter, focuses on the dynamic aspects of brain signals. It identifies temporal patterns that correlate with emotional responses, adapting over time to changes in the emotional state of the subject. 

To enhance information capture, we project \( x(t) \) using \( W_r^V \) as the value (representing Fuzzy Consequents), while the firing strength and membership degree are computed in the query space via the projection parameter \( W_r^Q \). Following TSK fuzzy model, the output ($Y_r(t)$) for rule \( r \) in these two proposed models is then expressed as:

\begin{equation}
Y_r(t) = \overline{f_{ML}}_{i,r}(W_r^Q x(t)) \cdot W_r^V x(t)
\end{equation}
where $Y_r(t)$ is the output of the penultimate layer of the network. Both $W_r^Q$ and $W_r^V$ are of dimension $D \times D$, where $D$ is the dimension of the input. To ameliorate issues with gradient descent, we add a $\ln$ operation after $Y_r(t)$, which stabilizes the gradient flow by computing logarithms of probabilities. 




Lastly, the Classifier, a single-layer linear network, links the outputs from the spatial and temporal filters to the output nodes. This configuration effectively integrates the processed signals into a structured format that is optimal for classification. 

The classifier employs a Cross-Entropy loss $\mathcal{L}_{CE}$ to measure the performance and guide the training process, which is defined as:

\begin{equation}
\mathcal{L}_{CE} = -\sum_{o=1}^N\sum_{c=1}^M y_{o,c} \log(p_{o,c}),
\end{equation}
where \( M \) is the number of classes, \( N \) is the number of samples, \( y_{o,c} \) is the ground truth label for class \( c \) of the \( o \)-th sample, and \( p_{o,c} \) is the predicted probability for class \( c \) of the \( o \)-th sample.

Together, these modules form a cohesive system that not only classifies emotional states with improved accuracy but also provides insights into the underlying neural mechanisms.

\section{Experiments and Results}
\label{sec:experiments&results}

\subsection{Dataset}

\subsubsection{fNIRS Dataset 1: Picture Recognition}

The first dataset involves 23 dyads of friends (\(age = 20.16 \pm 2.02\)) and 26 dyads of strangers (\(age = 19.47 \pm 2.21\)) participating in an image recognition task. Images were classified as neutral or negative, displayed for two seconds following a fixation point. The NIRScout 32\(\times\)32 was used to record blood oxygenation at a sampling rate of 7.8125 Hz, covering the prefrontal cortex with a 20-channel setup. Data preprocessing and additional details can be found in~\cite{chen2025interbrain} and~\cite{jiangFuzzybasedApproachPredict2024}. An example of these brain signals is shown in Fig.~\ref{fig:model_arch}(A).

\subsubsection{fNIRS Dataset 2: Picture Empathy}

This dataset examined responses to empathy-inducing images across 180 female participants divided into groups of 90 for social and physiological empathy, approved by the Henan Province Key Laboratory of Psychology and Behavior (20200702002). Each session included eight blocks alternating between negative and neutral images, structured in an ABBA sequence to control for order effects. Ratings of perceived pain were recorded on a 1 to 9 Likert scale. The experiment setup and data processing were consistent with the Picture Recognition dataset.

\subsubsection{EEG Dataset 3: FACED}

FACED involved eliciting emotional responses through 28 video clips(12 Negative videos, 12 Positive videos, and 4 Neutral videos). A total of 123 participants (average age 23.2 years) viewed videos varying in length while EEG data were collected. The analysis focused on the power spectral densities across five EEG frequency bands for 30-second epochs, generating 150 features per channel. Further methodological details are presented in ~\cite{chen2023large}.

\subsection{Comparative Analysis of other Models}

\begin{figure}[ht]
    \centering
    \includegraphics[width=0.9\linewidth]{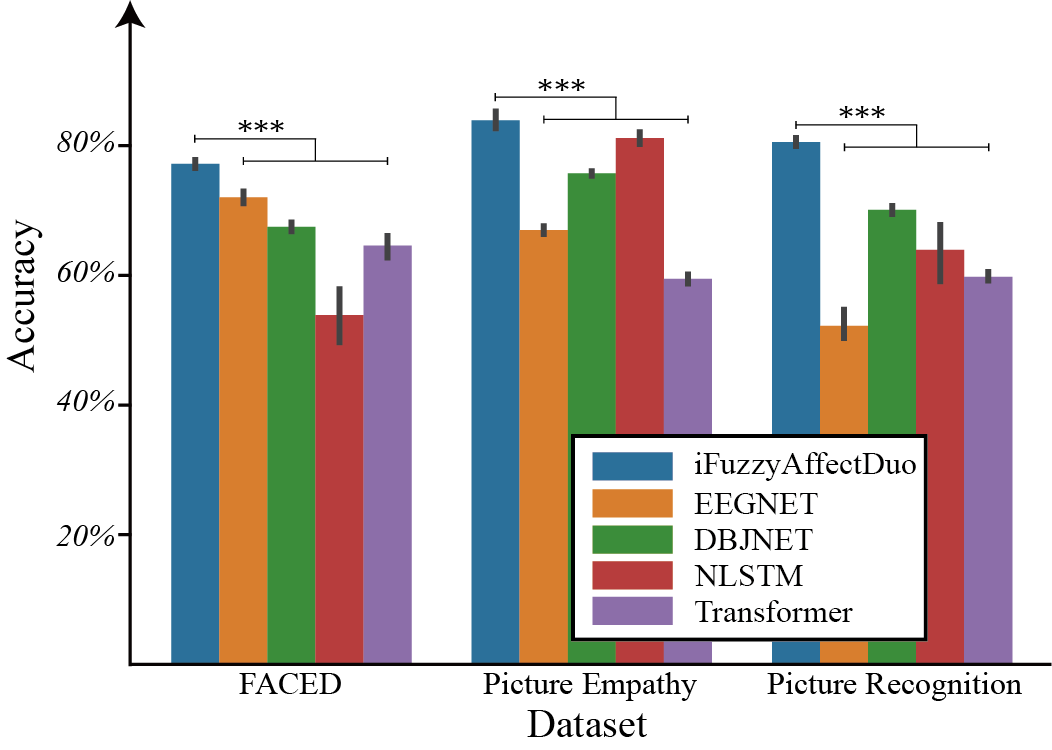}
    \caption{Comparative performance of the iFuzzyAffectDuo model against DBJNet, EEGNET, NLSTM, and Transformer models across three datasets. This figure demonstrates the model's robust performance superiority across diverse machine learning challenges. ***$p < 0.001$}
    \label{fig:model_performance}
\end{figure}

This section presents a comparative analysis of the iFuzzyAffectDuo model against existing models such as DBJNet, EEGNET, NLSTM, and Transformer across three datasets, employing paired t-tests with two-sided alternative hypotheses and the False Discovery Rate Benjamini/Hochberg (FDR-BH) correction for multiple comparisons. The results, visualized in Fig.~\ref{fig:model_performance}, show that the iFuzzyAffectDuo model consistently outperforms the others. On the Picture Recognition dataset, it achieved an accuracy of $80.53\% \pm 1.55\%$, significantly higher than the compared models with $p < 0.001$. Similar superiority is observed in the FACED dataset with an accuracy of $77.19\% \pm 1.49\%$, and in the Picture Empathy dataset with $83.88\% \pm 3.04\%$, both also surpassing competing models at $p < 0.001$ levels.

These findings illustrate the iFuzzyAffectDuo model's exceptional ability to manage and analyze diverse datasets, thereby asserting its versatility and effectiveness in complex machine learning landscapes, especially aBCI tasks. The consistently significant performance advantages across varied tasks not only validate the model's robust feature extraction and learning capabilities but also highlight its potential as a benchmark model in aBCI research.

\subsection{Comparison of Modified-Laplace and Gaussian Membership Functions}

In our comparative analysis using a model architecture with 5 rules, Modified-Laplace MFs consistently outperformed Gaussian MFs across various datasets. Specifically, within the FACED dataset, Modified-Laplace MFs achieved an accuracy of \(77.19\% \pm 1.49\%\), significantly higher than the \(74.55\% \pm 1.20\%\) observed for Gaussian MFs, with a notable statistical difference (\(t(20) = 6.55\), \(p < 0.001\), $Cohen's~d=1.95$). In the Picture Empathy dataset, Modified-Laplace MFs recorded \(83.88\% \pm 3.04\%\) accuracy, slightly outperforming the \(82.92\% \pm 3.01\%\) by Gaussian MFs (\(t(20) = 4.67\), \(p < 0.001\), $Cohen's~d=0.32$). Finally, for the Picture Recognition task, Modified-Laplace MFs demonstrated \(80.53\% \pm 1.55\%\) accuracy, surpassing the \(79.77\% \pm 2.11\%\) achieved by Gaussian MFs (\(t(20) = 2.75\), \(p < 0.05\), $Cohen's~d=0.36$). These findings underscore the effectiveness of Modified-Laplace MFs in enhancing classification accuracy across diverse settings.

\subsection{Fuzzy Set Membership and Feature Distribution Across EEG Signals}

\begin{figure*}
    \centering
    \includegraphics[width=0.9\linewidth]{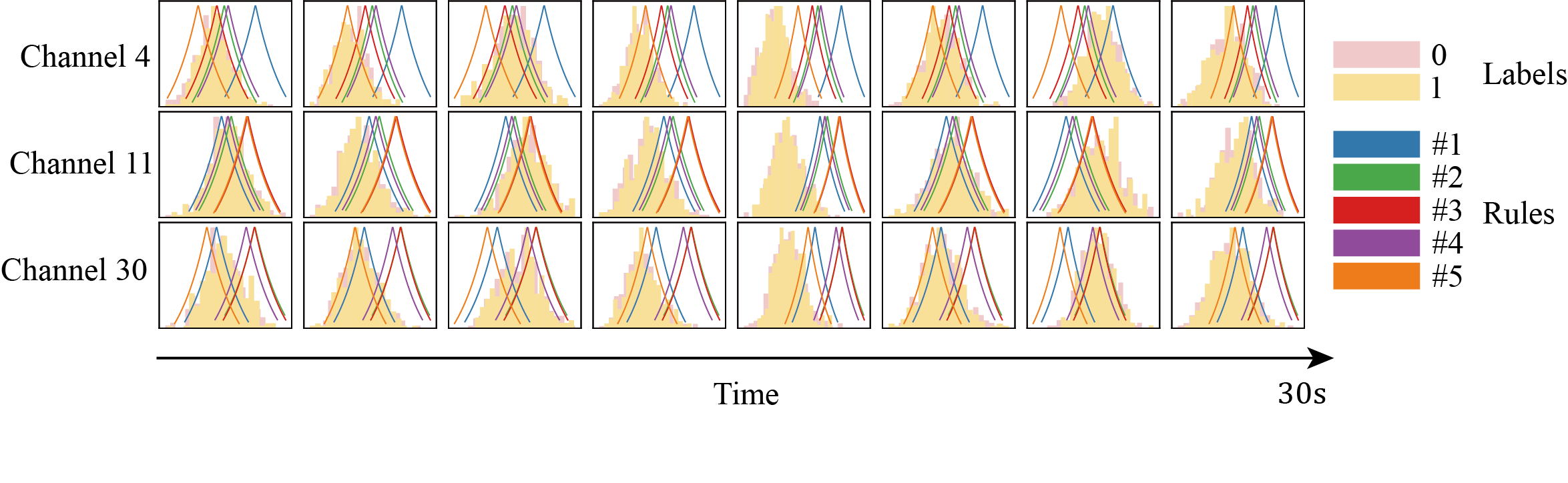}
    \caption{Dynamic Distribution of Fuzzy Set Membership and Feature Distributions Across the FACED Dataset at Key Time Intervals. This figure illustrates the variation in fuzzy membership degrees and overlays of neural response histograms. The histograms represent the EEG data in query space \(W_r^Q\), while the colored curves denote different Modified-Laplace MFs applied to decode underlying neural mechanisms.}
    \label{fig:feature_distribution}
\end{figure*}

The distributions of fuzzy set memberships depicted in Fig.~\ref{fig:feature_distribution} highlight the operational characteristics of Modified-Laplace MFs within the query space in FACED dataset. Notably, Channel 4 exhibits a broader diversity of rule centers (\(m\)), indicating a wide variation in feature representation. In contrast, Channel 11 demonstrates less diversity in rule centers, suggesting a more uniform feature response. Channel 30 reveals that while some rules appear similar, distinct patterns emerge, particularly between rules \#1 and \#5 versus rules \#2, 3, and 4, underscoring the subtle complexities in neural processing across different EEG channels.

\begin{table}[htbp]
\caption{Sample analysis: rule firing strength with top3 channel identified.}
\begin{center}
\renewcommand{\arraystretch}{0.8} 
\footnotesize 
\begin{tabular}{|c|l|l|l|}
\hline
\textbf{\centering Rule} & \multicolumn{3}{|c|}{\textbf{Top Channels with Strength}} \\
\cline{2-4} 
 & \textbf{\textit{Top1}} & \textbf{\textit{Top2}} & \textbf{\textit{Top3}} \\
\hline
\multicolumn{4}{|l|}{Picture: Neutral; Empathy: Physiological} \\
\hline
1 & 18: 0.26 & 11: 0.25 & 5: 0.19 \\
2 & 12: 0.18 & 1: 0.16 & 4: 0.10 \\
3 & 20: 0.96 & 6: 0.96 & 8: 0.87 \\
4 & 11: 0.24 & 18: 0.23 & 3: 0.15 \\
5 & 16: 0.94 & 15: 0.91 & 13: 0.87 \\
\hline
\multicolumn{4}{|l|}{Picture: Negative; Empathy: Physiological} \\
\hline
1 & 5: 0.29 & 4: 0.27 & 15: 0.26 \\
2 & 1: 0.13 & 8: 0.12 & 16: 0.12 \\
3 & 9: 0.91 & 20: 0.69 & 13: 0.62 \\
4 & 12: 0.39 & 6: 0.38 & 11: 0.38 \\
5 & 3: 0.85 & 17: 0.78 & 10: 0.72 \\
\hline
\multicolumn{4}{|l|}{Picture: Neutral; Empathy: Social} \\
\hline
1 & 18: 0.26 & 17: 0.20 & 15: 0.14 \\
2 & 16: 0.09 & 18: 0.09 & 5: 0.09 \\
3 & 3: 0.99 & 11: 0.98 & 19: 0.97 \\
4 & 2: 0.39 & 1: 0.37 & 18: 0.16 \\
5 & 9: 0.95 & 10: 0.87 & 15: 0.74 \\
\hline
\multicolumn{4}{|l|}{Picture: Negative; Empathy: Social} \\
\hline
1 & 18: 0.26 & 12: 0.26 & 14: 0.26 \\
2 & 15: 0.14 & 7: 0.14 & 13: 0.14 \\
3 & 9: 0.46 & 13: 0.41 & 10: 0.40 \\
4 & 7: 0.23 & 2: 0.20 & 14: 0.20 \\
5 & 3: 0.80 & 5: 0.80 & 8: 0.77 \\
\hline
\end{tabular}
\label{tab:sample_ana}
\end{center}
\end{table}

\subsection{Sample-wise Interpretability Analysis}

\begin{figure}[t]
  \centering
  \includegraphics[width=0.9\linewidth]{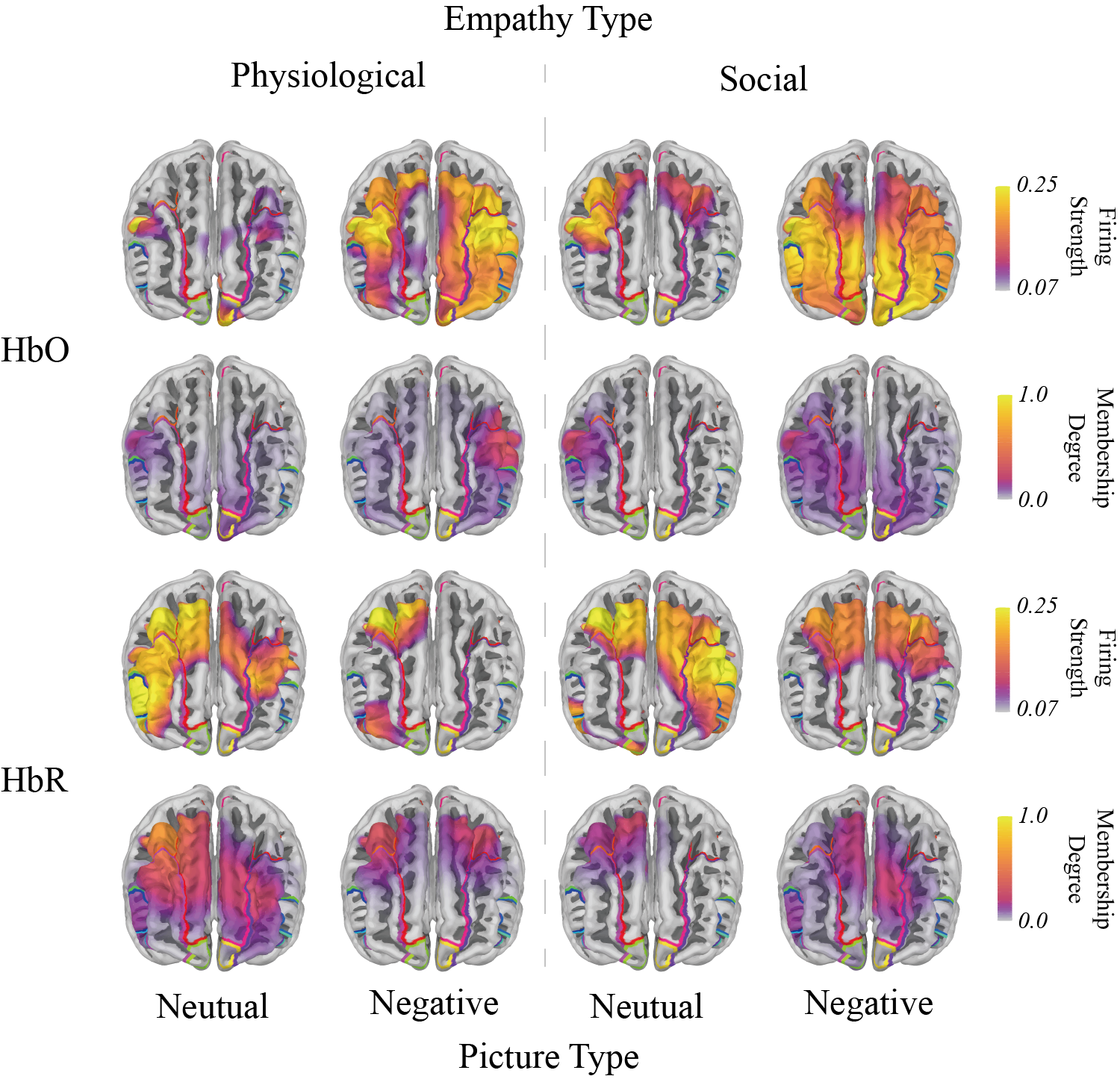}
  \caption{Sample analysis: Visualization of membership degrees and firing strengths in rule \#1.}
  \label{fig:case_analysis}
\end{figure}

In the Picture Empathy dataset, to provide an intuitive illustration of how the fuzzy filter discerns picture types and elucidates the neural patterns of different empathy types from fNIRS data, we present four demonstrative samples from the top-performing model. These samples represent a $2 \times 2$ factorial design (Picture type: Neutral or Negative; Empathy type: Physiological or Social). From Table~\ref{tab:sample_ana}, it is evident that Rule \#3 predominantly influences border firing strength in all four cases, while Rule \#2 consistently plays a minor role. The learned patterns of channels show substantial variety across different rules. Notably, significant contributions are observed from several specific channels: Channel 11, which measures oxy-hemoglobin (HbO) in the Left Frontopolar Area (FPA); Channel 18, which measures deoxy-hemoglobin (HbR) in the Right Ventrolateral Prefrontal Cortex (VLPFC); Channel 5, which measures HbO in the Left Ventromedial Prefrontal Cortex (VmPFC); and Channel 3, which measures HbO in the VLPFC. These channels demonstrate significant contributions across the rules. Particularly, in cases where the Picture is Neutral and the Empathy is Physiological, channel 16 by Rule \#5 and channel 20 by Rule \#3 are highlighted. Channel 11 is highlighted by Rules \#1 and \#4, and Channel 18 also receives significant emphasis by these rules. All membership degrees and firing strengths for these cases are depicted in Fig.~\ref{fig:case_analysis}. The demos for the decision-making processes in the Picture Recognition dataset are shown in Fig.~\ref{fig:model_arch} (E, F, G, and H). The orientation of the 3D Brain can be found in Fig.~\ref{fig:model_arch}(C).

Overall, these findings highlight the nuanced roles of the prefrontal cortex (PFC) in modulating distinct empathetic responses, with heightened activity observed in response to affective cues, illustrating the intricate interplay between neural circuitry and emotional processing~\cite{fusar2009functional,perry2017effects, Stuss2002AdultCN, Stuss2011FunctionsOT}.

\section{Conclusions}
\label{sec:conclusions}

This study introduces iFuzzyAffectDuo, a novel Fuzzy logic-based model for aBCI tasks, utilizing fNIRS and EEG technologies. It advances interpretability in neural decoding, outperforming Gaussian MFs, CNN-based, and self-attention models in accuracy across two fNIRS and one EEG dataset. The Modified-Laplace distribution enhances the model's ability to detail neural activity patterns, making complex decisions understandable. The potential of iFuzzyAffectDuo extends to Functional Magnetic Resonance Imaging (fMRI) and Electrocorticography (ECoG) applications, potentially enriching theoretical frameworks in neuroscience. Future work will address the model's hyperparameter sensitivity and training duration, explore online testing, and integrate multi-modal technologies to broaden its application.

\bibliographystyle{IEEEtran}
\bibliography{ref}
\end{document}